\documentstyle[12pt]{article}\textheight 230mm\textwidth 150mm
            \newcommand{\be}{\begin{eqnarray}}
            \newcommand{\ee}{\end{eqnarray}}
           \newcommand{\eel}[1]{\label{#1}\end{eqnarray}}
\newcommand{\e}[1]{\label{e:#1}\end{eqnarray}}
     \newcommand{\eg}{{\em e.g.\ }}
            \newcommand{\ie}{{\em i.e.\ }}

  \newcommand{\La}{{\Lambda}}
            
            \newcommand{\del}{{\delta}}
 \newcommand{\om}{{\omega}}

            \newcommand{\beq}{\begin{quote}}
            \newcommand{\eq}{\end{quote}}
            
            \newcommand{\al}{\alpha}
            \newcommand{\ben}{\begin{enumerate}}
            \newcommand{\een}{\end{enumerate}}
            \newcommand{\bit}{\begin{itemize}}
            \newcommand{\ei}{\end{itemize}}
    	\newcommand{\nn}{\nonumber}
            \newcommand{\r}[1]{(\ref{e:#1})}
            \newcommand{\edfl}[1]{\Label{#1}\end{df}}

\newcommand{\ve}{{\varepsilon}}

\newcommand{\dif}{{\partial}}
\newcommand{\half}{\frac{1}{2}}
	
	\newcommand{\ldif}{{\stackrel{\leftarrow}{\partial}}}
\newcommand{\ldel}{{\stackrel{\leftarrow}{\delta}}}

\begin{document}
\begin{titlepage}

\vspace*{40mm}
\begin{center}{\LARGE\bf
Generalized Poisson sigma models.}\end{center}
\vspace*{3 mm}
\begin{center}
\vspace*{3 mm}

\begin{center}Igor Batalin\footnote{On leave of absence from
P.N.Lebedev Physical Institute, 117924  Moscow, Russia\\E-mail:
batalin@td.lpi.ac.ru.} and Robert
Marnelius\footnote{E-mail: tferm@fy.chalmers.se}
 \\ \vspace*{7 mm} {\sl
Institute of Theoretical Physics\\ Chalmers University of Technology\\
G\"{o}teborg University\\
S-412 96  G\"{o}teborg, Sweden}\end{center}
\vspace*{25 mm}
\begin{abstract}
A general master action in terms of superfields is given which  generates
generalized Poisson sigma models by means of a natural
ghost number prescription. The simplest
representation is the sigma model considered by Cattaneo and Felder. For Dirac
brackets considerably more general models are generated.
\end{abstract}\end{center}\end{titlepage}
Poisson sigma models \cite{Ikeda,SS} are presently of general interest after the appearance of
Kontsevich's \cite{Kon} general formula for deformation quantization and its
Poisson
sigma model formulation of Cattaneo and Felder \cite{CF,CF1}.
A unique geometrical description of sigma models was given in \cite{AKSZ}.
 In this
communication we present a new interpretation of this model and at the same time
generalize it considerably. Our formulation starts from a master action in a
manifestly supersymmetric form. Our master action is
\be
&&\Sigma[\Phi, \Phi^*]=\int d^2u d^2\tau
\left(\Phi^*_A(u,\tau)D\Phi^A(u,\tau)(-1)^{\ve_A}-S(\Phi(u,\tau),\Phi^*(u,
\tau)\right),
\e{1}
where $u^a$, $a=1,2$ are bosonic coordinates on a disc and $\tau^a$, $a=1,2$,
corresponding fermionic ones, \ie $u^a$ are even and $\tau^a$ odd. The
superfields $\Phi^A(u,\tau)$ and their corresponding
super-antifields  $\Phi^*_A(u,\tau)$ have Grassmann parities $\ve_A$ and
$\ve_A+1$
respectively. The de Rham
differential $D$  is as in
\cite{CF} given by
\be
&&D\equiv \tau^a{\dif\over\dif u^a}, \quad \Rightarrow
\quad D^2=0.
\e{2}
  In order for $\Sigma$
to be a master action we need to prescribe some properties of $S$. If we view
$\Sigma$ as an ordinary action then the equations of motion from \r{1} become\be
&&D\Phi^A=(S, \Phi^A),\quad D\Phi^*_A=(S, \Phi^*_A),
\e{3}
where we have introduced a local antibracket defined by
\be
&&(f, g)\equiv
f{\stackrel{\leftarrow}{\dif}\over\dif\Phi^A}{\stackrel{\rightarrow}{\dif}
\over\dif\Phi^*_A}g-(f\leftrightarrow
g)(-1)^{(\ve(f)+1)(\ve(g)+1)}.
\e{4}
Equations \r{3} imply
\be
&&0=D^2\Phi^A=(S,(S,\Phi^A))=(\half(S,S), \Phi^A),\nn\\&&
0=D^2\Phi^*_A=(S, (S, \Phi^*_A))=(\half(S,S), \Phi^*_A),
\e{401}
and
\be
&&DS=(S,S),\quad D^2S=(S,(S, S))\equiv 0.
\e{402}
Consistency of \r{401} requires
\be
&&(S,S)=0,
\e{5}
which when inserted into \r{402}  implies that $S$ satisfies the equation
$DS=0$.
We impose the condition \r{5} on the   local master action $S$. In order to
determine
its general form we have also to impose a ghost number prescription. We
choose the odd coordinates
$\tau^a$ to have ghost number plus one. This implies that $D$ has ghost
number plus
one. $D$ could then be interpreted to be  a kind of BRST charge operator
in which
case the action \r{1} has a similar structure to the master action for the
(closed)
string field theory by Zwiebach \cite{Zwie}. Since we require $\Sigma$ to have ghost
number zero and since the measure $d^2\tau$ has ghost number minus two we
are led to
the following general rule for the superfields and
$S$:
\be
&&gh(\Phi^A)+gh(\Phi^*_A)=1, \quad gh(S)=2.
\e{6}
Such a ghost prescription was previously  considered in \cite{BT} for a master
action in connection with the Dirac bracket.

Before we analyze the general form of $S$ that follows from the ghost
prescription
above, we will show that the action \r{1} under the condition \r{5} also
satisfies a
master equation. Let us first define functional derivatives of the
superfields by
\be
&&{\del\over\del\Phi^B(u',\tau')}\Phi^A(u,\tau)=\del^A_B\del^2(u-u')\del^2(
\tau-\tau'),\nn\\
&&{\del\over\del\Phi^*_B(u',\tau')}\Phi^*_A(u,\tau)=\del^B_A\del^2(u-u')\del
^2(\tau-\tau'),
\e{7} where
\be
&&\del^2(\tau-\tau')\equiv\half\ve_{ab}(\tau^b-\tau^{\prime
b})(\tau^a-\tau^{\prime
a}),
\e{8}
where in turn we use the convention
\be
&&\ve_{ab}=-\ve_{ba},\quad \ve^{ab}=-\ve^{ba}, \quad
\ve^{ab}\ve_{bc}=\del^a_c,\quad
\ve^{12}=-\ve_{12}=1.
\e{9}
We choose the integration convention
\be
&&\int
d^2\tau\tau^a\tau^b=\ve^{ab}
\e{10}
in terms of which \r{8} is an appropriate delta function.
In terms of the functional derivatives \r{7} we may now define a functional
antibracket by
\be
&&(F, G)'\equiv \int
F{\ldel\over\del\Phi^A(u,\tau)} d^2u
d^2\tau{\stackrel{\rightarrow}{\del}\over\del\Phi^*_A(u,\tau)}G-(F\leftrightarrow
G)(-1)^{(\ve(F)+1)(\ve(G)+1)},\nn\\
\e{11}
as well as a related nilpotent $\Delta$-operator
\be
&&\Delta'\equiv\int d^2u d^2\tau
(-1)^{\ve_A}{\del\over\del\Phi^A(u,\tau)}{\del\over\del\Phi^*_A(u,\tau)}.
\e{12}
From \r{1} and \r{11} we find then
\be
\half(\Sigma, \Sigma)'=\int d^2u
d^2\tau\left((-1)^{\ve_A}D\Phi^*_AD\Phi^A-DS+\half(S,S)\right)=0,
\e{13}
since the first term is zero by partial integration, the second term is
zero directly
and the third is zero due to our condition \r{5}. Furthermore, we have
\be
&&\Delta'\Sigma=0
\e{14}
since the $\tau$-part yields a factor zero. (As usual we believe that the
bosonic part may be regularized appropriately.) $\Sigma$ satisfies
therefore also the
quantum master equation
\be
&&\half(\Sigma, \Sigma)'=i\hbar\Delta'\Sigma,
\e{15}
which means that the quantization requires no quantum corrections to the measure.

The superfields above may be decomposed into ordinary fields in such a
fashion that
the antibracket \r{11} and the $\Delta$-operator \r{12} have the
conventional forms.
We may \eg define the components as follows (notice that the component
fields have
different ghost numbers):
\be
&&\Phi^A(u,\tau)=\Phi^{0A}(u)+\tau^a\ve_{ab}\Phi^{bA}(u)+
\half\ve_{ab}\tau^b\tau^a\Phi^{3A}(u),\nn\\
&&\Phi^*_A(u,\tau)=\Phi^*_{3A}(u)+
\Phi^*_{aA}(u)\tau^a+\half\ve_{ab}\tau^b\tau^a\Phi^*_{0A}(u).
\e{16}
The solutions to \r{7}  are then
\be
&&{\del\over\del\Phi^A(u,\tau)}=
{\del\over\del\Phi^{3A}(u)}+(-1)^{\ve_A}\tau^a{\del\over\del\Phi^{aA}(u)}
+\half\ve_{ab}\tau^b\tau^a{\del\over\del\Phi^{0A}(u)}\nn\\
&&{\del\over\del\Phi^*_A(u,\tau)}={\del\over\del\Phi^*_{0A}(u)}+\tau^a\ve_{ab}
{\del\over\del\Phi^*_{bA}(u)}+\half\ve_{ab}\tau^b\tau^a{\del\over\del\Phi^*_
{3A}(u)}.
\e{17}
These properties imply then that $\Phi^*_{nA}(u)$ become  antifields to
$\Phi^{nA}(u)$, ($n=0,1,2,3$). We have from \r{11} and \r{12}
\be
&&(F, G)'= \int
F{\ldel\over\del\Phi^{nA}(u)} d^2u
{\stackrel{\rightarrow}{\del}\over\del\Phi^*_{nA}(u)}G-(F\leftrightarrow
G)(-1)^{(\ve(F)+1)(\ve(G)+1)},\nn\\
&&\Delta'\equiv\int d^2u
(-1)^{\ve_{nA}}{\del\over\del\Phi^{nA}(u)}{\del\over\del\Phi^*_{nA}(u)},\quad
\ve_{nA}\equiv\ve(\Phi^{nA}).
\e{171}
One may easily check that this expression for the antibracket together with
\r{16} implies
\be
&&(\Phi^A(u,\tau), \Phi^*_B(u',\tau'))'=\del^A_B\del^2(u-u')\del^2(\tau-\tau'),
\e{172}
which trivially follows from \r{11}.
\\

\noindent
{\bf Examples}:

A generalized Poisson sigma model is obtained if we as superfields $\Phi^A$
choose
$X^i(u,\tau)$, ($\ve(X^i)\equiv\ve_i$), and corresponding antifields
$X^*_i(u,\tau)$,
($\ve(X^*_i)=\ve_i+1$) with ghost numbers zero and plus one respectively. A
local
master action $S$ with ghost number plus two must then be quadratic in
$X^*_i$. A
general ansatz is
\be
&&S(X,X^*)=\half X^*_iX^*_j\om^{ji}(X)(-1)^{\ve_i}.
\e{18}
The master equation \r{5} or \r{13} yields then
\be
&&\om^{ij}(X)\ldif_l\om^{lk}(X)(-1)^{\ve_i\ve_k}+cycle(ijk)=0.
\e{181}
Since \r{18} requires $\om^{ji}(X)$ to have the symmetry property
\be
&&\om^{ij}(X)=-\om^{ji}(X)(-1)^{\ve_i\ve_j}
\e{1811}
we may identify $\om^{ji}(X)$ with the (super) Poisson bracket $\{X^j, X^i\}$.
(Eqs.\r{181} are their Jacobi identities.)   The
master action
$\Sigma[X,X^*]$ is therefore the master action for  a Poisson sigma model and it
agrees with the one given by Cattaneo and Felder
\cite{CF} for
$\ve_i=0$ and with the identification
$\eta_i=X^*_i$. The original fields are the component fields with ghost
number zero.

In a theory with constraints
\be
&&\theta^{\al}(X)=0,
\e{183}
we define the following local master action
\be
&&S(X,\La;X^*,\La^*)=\half
X^*_iX^*_j\om^{ji}(X)(-1)^{\ve_i}+\La^*_{\al}\theta^{\al}(X),
\e{19}
where we have introduced new superfields $\La^{\al}(u,\tau)$ and
$\La^*_{\al}(u,\tau)$
with the Grassmann parities $\ve_{\al}+1$ and
$\ve_{\al}\equiv \ve(\theta^{\al})$.
The ghost numbers for  $\La^{\al}$ and
$\La^*_{\al}$ must be  minus
one and plus two
 respectively according to the rule \r{6}.
The master equation for \r{19} yields
the Jacobi identities \r{181} and the degeneracy conditions
\be
&&\theta^{\al}(X)\ldif_j\om^{ji}(X)=0.
\e{182}
This means that one may identify $\om^{ji}(X)$ with
the (super) Dirac bracket in a constraint theory \cite{Dirac}.
 The expression \r{19}
for
$S$ means also that there is a larger, reducible gauge invariance for Dirac
brackets
on the constraint surface as compared to the case
\r{18}. With the same superfields we may define even more general
expressions for $S$
(cf eq.(4.27) in \cite{BT}):
\be
&&S(X,\La;X^*,\La^*)=\half
X^*_iX^*_j\om^{ji}(X)(-1)^{\ve_i}+\La^*_{\al}\theta^{\al}(X)+\nn\\&&+
(-1)^{\ve_j+\ve_i\ve_k}{1\over
6}X^*_iX^*_jX^*_k\om^{kji}_{\al}(X)\La^{\al}+
(-1)^{\ve_i}X^*_i\La^*_{\al}\om^{\al
i}_{\beta}(X)\La^{\beta}+\ldots,
\e{20}
where $\om^{kji}_{\al}$ is totally (super) antisymmetric in $i,j,k$ with
Grassmann
parity $\ve_{ijk}\equiv\ve_i\ve_j+\ve_j\ve_k+\ve_k\ve_i$.
 Nonzero $\om^{\al
i}_{\beta}(X)$ and $\om^{kji}_{\al}(X)$ imply that the master equation
yields a weak form of the degeneracy conditions \r{182} and the Jacobi
identities
\r{181}:
\be
&&\theta^{\al}(X)\ldif_j\om^{ji}(X)=\om_{\beta}^{\al i}(X)\theta^{\beta}(X)(-1)^{\ve_i},\nn\\
&&\om^{ij}(X)\ldif_l\om^{lk}(X)(-1)^{\ve_i\ve_k}+cycle(ijk)=\om_{\al}^{ijk}(X)\theta^{\al}(X).
\e{201}
 This means that
$\om^{ji}(X)$ here is a generalized (weak) Dirac bracket. A still more
general form of
$S$ is obtained if we introduce new superfields $\Xi^{\al_1}$ and
$\Xi^*_{\al_1}$ with ghost numbers
minus two and plus three respectively. We may then add to $S$
the following $\Xi$-dependent terms
\be
&&\Xi^*_{\al_1}Z^{\al_1}_{\al}(X)\La^{\al}+(-1)^{\ve_i}X^*_i\Xi^*_{\al_1}\om
^{\al_1
i}_{\beta_1}(X)\Xi^{\beta_1}+(-1)^{\ve_{\al}+1}\half\La^*_{\al}\La^*_{\beta}
\om^{\beta\al}_{\beta_1}(X)\Xi^{\beta_1}+\nn\\&&+(-1)^{\ve_i}\half X^*_i\Xi^*_{\al_1}\om_{\beta\al}^{\al_1 i}(X)\La^{\al}\La^{\beta}(-1)^{\ve_{\beta}}+\ldots.
\e{21}
The master equation yields then \eg
\be
&&Z_{\al}^{\al_1}(X)\theta^{\al}(X)=0,
\e{211}
\be
&&(-1)^{\ve_{\al}\ve_j}Z^{\al_1}_{\al}(X)\ldif_i\om^{ij}(X)-\om_{\beta_1}^{\al_1 j}(X)Z^{\beta_1}_{\al}(X)+\nn\\&&+Z^{\al_1}_{\beta}(X)\om_{\al}^{\beta j}(X)(-1)^{\ve_j}+\om_{\al\beta}^{\al_1 j}(X)\theta^{\beta}(X)=0.
\e{212}
The  relation \r{211}  means that the constraints are reducible (linearly dependent). The last relation multiplied by $\theta^{\al}$ from the right yields the first relation in \r{201} by means of \r{211}. $Z_{\al}^{\al_1}$ are assumed to be linearly independent (first stage reducibility)  and the Grassmann parities are $\ve(\Xi^{\al_1})=\ve_{\al_1}$, $\ve(\Xi_{\al_1}^*)=\ve_{\al_1}+1$ for $\ve(Z_{\al}^{\al_1})=\ve_{\al}+\ve_{\al_1}$. If $Z_{\al}^{\al_1}$ are linearly dependent (higher stage reducibility) then we need new field-antifield pairs to enter the master action.\\

\noindent
{\bf Generalizations}:

Our general superfield formulation may be further generalized. First one
may notice
that all formal properties are valid for any even measure. We may therefore
consider  master actions defined on  a surface of any even dimension. However,
since the ghost number prescriptions depends on this dimension  different kinds
of models will then be generated. (Our formulas do not apply to odd dimensions. Compare the one dimensional superfield treatment in \cite{GD}.)

We may choose arbitrary coordinates $u^a$ on the surface. For the measure
\be
&&d^2u\left(\det h^b_a(u)\right)^{-1}
\e{2121}
we have the $D$-operator
\be
&&D=\tau^aT_a+\half\tau^b\tau^aU_{ab}^c(u){\dif\over\dif\tau^c},\quad T_a\equiv h_a^b(u){\dif\over\dif u^b},
\e{2122}
which is nilpotent for
\be
&&U_{ab}^c(u)=-h_a^f(u)h_b^d(u){\dif\over\dif u^f}(h^{-1})_d^c(u)-(a\leftrightarrow b),
\e{2124}
since
\be
&&D^2=0\;\Leftrightarrow\;[T_a, T_b]=U_{ab}^c(u)T_c.
\e{2123}

One may also generalize the formalism along the lines of the generalized
antisymplectic formulation in
\cite{BT1}. Let $Z^I(u,\tau)$, ($\ve(Z^I)\equiv\ve_I$), denote general
antisymplectic
superfields. Our master action may then be written as
\be
&&\Sigma[Z]=\int d^2u
d^2\tau\left(V_I(Z(u,\tau))DZ^I(u,\tau)(-1)^{\ve_I}-S(Z(u,\tau))\right),
\e{22}
where the antisymplectic potential $V_I$ has the Grassmann parity
$\ve(V_I)=\ve_I+1$.
The equations of motion are
\be
&&E_{KI}(Z)DZ^I(-1)^{\ve_I}-\dif_KS=0\quad \Leftrightarrow\quad DZ^I=(S,Z^I),
\e{23}
where
\be
&&E_{KI}(Z)\equiv \dif_KV_I-\dif_IV_K(-1)^{\ve_K\ve_I}, \quad
(f,g)=f\ldif_IE^{IJ}\stackrel{\rightarrow}{\dif}_Jg,\quad
E^{IJ}E_{JK}=\del^I_K.\nn\\
\e{24}
As before consistency requires $(S,S)=0$ and also that $S$ is determined by
the ghost
number prescription $gh(S)=2$ and
\be
&&gh(V_I)=1-gh(Z^I)\quad\Leftrightarrow\quad gh(E_{KI})=1-gh(Z^K)-gh(Z^I).
\e{25}
A still more general form for $\Sigma$ is
\be
&&\Sigma[Z]=\int d^2u
d^2\tau\left(Z^K(u,\tau)\bar{E}_{KI}(Z(u,\tau))DZ^I(u,\tau)(-1)^{\ve_I}-S(Z(
u,\tau))\right),
\e{26}
where
\be
&&\bar{E}_{KI}(Z)\equiv (Z^J\dif_J+2)^{-1}E_{KI}(Z)=\int_0^1d\al\al
E_{KI}(\al Z).
\e{27}
The corresponding expression for the symplectic case was given in \cite{BF}.

The above master actions $\Sigma[Z]$ satisfy the master equation
\be
&&(\Sigma, \Sigma)'=\int d^2u d^2\tau \Sigma{\ldel\over\del
Z^I(u,\tau)}E^{IK}(Z(u,\tau)){\stackrel{\rightarrow}{\del}\over\del
Z^K(u,\tau)}\Sigma=0,
\e{28}
from which it follows that $\Sigma$ is gauge invariant under nilpotent gauge
generators. We have
\be
&&\int R_I^{\;K}(u,\tau;u',\tau')d^2u'd^2\tau'{{\del}\over\del
Z^K(u',\tau')}\Sigma=0,\nn\\
&&\left.\int R_I^{\;K}(u,\tau;u',\tau') d^2u'd^2\tau'
R_K^{\;J}(u',\tau';u'',\tau'')\right|_{{\del\Sigma/\del
Z}=0}=0,
\e{29}
where
\be
&&
R_I^{\;K}(u,\tau;u',\tau')\equiv
\biggl[2\left({\stackrel{\rightarrow}{\del}\over\del
Z^I(u,\tau)}\Sigma{\ldel\over\del
Z^J(u',\tau')}\right)+\nn\\&&+\Sigma{\stackrel{\rightarrow}{\del}\over\del
Z^I(u,\tau)}{\ldel\over\del Z^J(u',\tau')}\biggr]E^{JK}(Z(u',\tau')).\nn\\
\e{30}
\\

\noindent
{\bf Acknowledgements}:

I.A.B. would like to thank Lars Brink for
his very warm hospitality at the
Department of Theoretical Physics, Chalmers
and G\"oteborg University.
 The work of I.A.B. is supported by the grants 99-01-00980 and
99-02-17916 from Russian foundation for basic researches and by the
President grant
00-15-96566 for supporting leading scientific schools. The work is partially supported by INTAS grant 00-00262.

\end{document}